\documentclass{article}

\usepackage[pdftex]{graphicx}
\graphicspath{{./graphics/}}
\DeclareGraphicsExtensions{.pdf}

\usepackage{amsmath}
\usepackage{amssymb}
\usepackage{algorithmic}
\usepackage{array}
\usepackage{color}
\usepackage[hidelinks]{hyperref}
\usepackage[tableposition=top]{caption}
\usepackage{subcaption}
\usepackage{cite}
\usepackage[table,xcdraw]{xcolor}
\usepackage{spconf}
\usepackage{multirow}
\usepackage[symbol]{footmisc}

\usepackage{enumitem}
\setlist[itemize]{leftmargin=*}
\setlist[enumerate]{leftmargin=*}

\newlength{\bibitemsep}
\newlength{\bibparskip}
\let\oldthebibliography\thebibliography
\renewcommand\thebibliography[1]{%
  \oldthebibliography{#1}%
  \setlength{\parskip}{\bibitemsep}%
  \setlength{\itemsep}{\bibparskip}%
}

\begin{document}

\title{\vspace{-4mm}
Quantitative evidence on overlooked aspects of enrollment speaker embeddings for target speaker separation
}

\name{Xiaoyu~Liu \quad Xu~Li \quad Joan~Serr\`{a}}

\address{Dolby Laboratories}

\markboth{Submitted to ICASSP 2023}%
{Shell \MakeLowercase{\textit{et al.}}: Title}%
\maketitle
\ninept

\begin{abstract}
Single channel target speaker separation (TSS) aims at extracting a speaker's voice from a mixture of multiple talkers given an enrollment utterance of that speaker. A typical deep learning TSS framework consists of an upstream model that obtains enrollment speaker embeddings and a downstream model that performs the separation conditioned on the embeddings. In this paper, we look into several important but overlooked aspects of the enrollment embeddings, including the suitability of the widely used speaker identification embeddings, the introduction of the log-mel filterbank and self-supervised embeddings, and the embeddings' cross-dataset generalization capability. Our results show that the speaker identification embeddings could lose relevant information due to a sub-optimal metric, training objective, or common pre-processing. In contrast, both the filterbank and the self-supervised embeddings preserve the integrity of the speaker information, but the former consistently outperforms the latter in a cross-dataset evaluation. The competitive separation and generalization performance of the previously overlooked filterbank embedding is consistent across our study, which calls for future research on better upstream features.

\end{abstract}
%

\section{Introduction}
\label{sec:introduction}

Single channel target speaker separation (TSS) is the task of separating a speaker's voice from interfering talkers given a pre-recorded enrollment utterance that characterizes that speaker (the target speaker). A deep learning-based TSS framework typically consists of an upstream speaker embedding model and a downstream separation model, with the latter conditioned on the enrollment embeddings from the former, acting as target speaker references. For the upstream, in general, existing research performs two choices: (i) to use utterance- or frame-level embeddings, and (ii) to pre-train the embedding model as a separate module or fine-tune the embeddings together with the downstream. For utterance-level embeddings, many systems employ speaker identification (SID) networks pre-trained for a low equal error rate (EER) to extract a summary vector from the entire enrollment utterance~\cite{vfilter, percepnet, e3net, pdccrn, two-stage, atss-net, one-enhance-all}. On the other hand, frame-level embeddings take advantage of attention algorithms to align each mixture frame with the most informative enrollment frames~\cite{speaker-inventory, attention-scaling, speaker-aware}. For both (i) and (ii), fine-tuning outperforms pre-trained embeddings, based on the assumption that joint training captures more relevant speaker information for TSS~\cite{speaker-aware, time-speakerbeam}.

Despite of much research along those choices, there are still important but overlooked aspects of enrollment embeddings that require further attention. For the widely used SID embeddings, the effects of a low EER, commonly used pre-processing, and data augmentation on the TSS quality are unclear. In addition, we look at two new embeddings not explored for TSS before: log-mel filterbank (FBANK) and self-supervised learning (SSL). FBANK, as a simple signal processing method, has been ignored as an enrollment option in previous literature. SSL are a class of powerful models that learn problem-agnostic speech features from unlabelled data~\cite{pascual2019learning, wav2vec2, hubert}, and we hypothesize that such broader information (compared to SID) could benefit TSS enrollment. Note that, unlike~\cite{ssl-separation}, which uses SSL as the input mixture features for blind speaker separation, we limit SSL to offline processing the enrollment utterance, since TSS often requires real-time low-complexity processing for the mixtures~\cite{percepnet, e3net, pdccrn, two-stage}. Finally, we consider a cross-dataset evaluation to assess the generalization of the enrollment embeddings~\cite{kadiouglu2020empirical}, which is another important but overlooked aspect in previous TSS research.

Our work studies pre-trained utterance- and frame-level embeddings, as well as fine-tuned frame-level embeddings. Under each category, FBANK, SID, and SSL features are investigated in detail. Specifically, we provide answers to the following open questions:  

\begin{itemize}

\item \textbf{Does a lower EER mean better separation}? An upstream SID network with a low EER is a natural choice for TSS~\cite{two-stage, percepnet, speaker-aware}, but we show that EER is an unreliable metric for the success of TSS.

\item \textbf{Does feature normalization (FN) in SID improve TSS}? FN, a common pre-processing in SID\cite{zeinali2019but, ecapa-tdnn, xvector, zhao2020improving, zheng2006comparative}, normalizes the recording channel characteristics by subtracting the per-band mean from the input FBANK to the SID systems. We show that FN hurts TSS.

\item \textbf{Does data augmentation in SID improve TSS}? Augmenting training data by more speakers and distortions often benefits SID~\cite{xvector}, but we find that such embeddings may not benefit TSS much under both clean and noisy enrollment conditions.

\item \textbf{Which one is better, SSL or SID embeddings}? SSL encodes more speech information than SID, but we show that, to take advantage of SSL, frame-level embeddings are preferred over utterance-level ones, and that the studied SID embeddings do not benefit from frame-level information. 

\item \textbf{Does a more powerful SSL model yield better TSS}? By comparing two SSL models, we find that the more powerful one performs only marginally better. 

\item \textbf{Are fine-tuned embeddings better than the pre-trained ones}? To answer this question, we fine-tune the pre-trained SSL and SID models and only observe improvements by the SSL model.

\item \textbf{How does each embedding compare with FBANK}? Remarkably, the performance of the simple FBANK is close to or in some cases better than other studied embeddings.

\item \textbf{How generalizable are the embeddings to different test sets}? We show that FBANK generalizes competitively among various upstream features. We also observe that the pre-trained and fine-tuned SSL features could suffer from overfitting.  
\end{itemize}

With this extensive study, we hope to provide insight and a practical guide on speaker embeddings for TSS enrollment.


\section{Methodology}
\label{sec:method}

\subsection{Upstream enrollment embeddings}
\label{sec:upstream}

As mentioned, we consider pre-trained utterance- and frame-level embeddings, as well as fine-tuned frame-level embeddings. Under each of them, we look into FBANK, SID, and SSL features. First, we describe the baseline FBANK and the pre-trained utterance-level SID embeddings. Note that, in the following, SID models without `FN' in their names do not use feature normalization, and that models without `aug' in their names are trained on VoxCeleb1,2~\cite{voxceleb1, voxceleb2} with 7,205 speakers without data augmentation.

\begin{itemize}
\item \textbf{FBANK}: 80-dim FBANK features are computed for each 25\,ms frame with a hop size of 10\,ms between adjacent frames. The per-band temporal mean and standard deviation are concatenated to form a 160-dim utterance-level embedding vector.

\item \textbf{d-vector, d-vector-FN, d-vector-aug}: These d-vector models process the frame-level FBANK features by a 3-layer LSTM, each with 768~cells as in~\cite{dvector}, and generate a 256-dim unit-length embedding vector from the last time step. The d-vectors are trained with the cross-entropy loss. For FN, the per-band utterance mean is subtracted from the input FBANK. The d-vector-aug increases the training speakers to 18,470 by adding an in-house collection of data obtained from OpenSLR. We also augment the training data with SpecAugment~\cite{specaugment}, pitch shifting, as well as noise and reverberation from~\cite{dns-challenge}.

\item \textbf{e-vector, e-vector-FN}: The e-vectors refer to the ECAPA-TDNN model~\cite{ecapa-tdnn}, and they process frame-level FBANK features and yield a 256-dim embedding from a temporal pooling layer. Compared with d-vectors, e-vectors employ advanced TDNN-like blocks~\cite{ecapa-tdnn} and a margin-based AAM-softmax loss~\cite{arcface} for optimal clustering in the embedding space. Therefore, e-vectors yield state-of-the-art EER. We adopt the implementation from~\cite{speechbrain}. 
\end{itemize}

As for the pre-trained utterance-level SSL embeddings, we consider the following two models:

\begin{itemize}
\item \textbf{PASE+}: PASE+\cite{pascual2019learning} encodes a waveform into a sequence of 256-dim features, which are trained to predict multiple self-supervised objectives, including SID, FBANK, prosody, etc. We trained a PASE+ on the 960~hours of the LibriSpeech training set~\cite{librispeech}, and verified the model by training a d-vector with PASE+ features and achieving lower EER than the FBANK d-vector. An utterance embedding is obtained by simple average (we also tried concatenating the standard deviation but found no significant benefits). 

\item \textbf{HuBERT}: HuBERT~\cite{hubert} is a powerful model that relies on transformers to predict masked sound units. We take the official HuBERT-Large model from~\cite{superb}, which is pre-trained on the 60\,k-hour Libri-Light dataset~\cite{libri-light}. The model returns 1024-dim frame embeddings from all 25~transformer layers. Following~\cite{superb}, a temporal average on each layer obtains 25~single vectors, which are then merged by a learnable weighted sum in the downstream training (note that, in our work, the pre-trained HuBERT model is always frozen).
\end{itemize}

As for the pre-trained frame-level embeddings, we consider FBANK, d-vector, PASE+, and HuBERT. The d-vector outputs hidden states from all time steps at test time, but during pre-training it uses the last time step in the cross-entropy loss. The weighted sum in HuBERT reuses the learned coefficients obtained from the utterance-level experiments. To study the jointly optimized frame-level embeddings, all the weights in the pre-trained PASE+ and d-vector models are also fine-tuned in the downstream training stage to optimize the separation loss of the corresponding model.

\begin{table*}[t]
\centering
\vspace{-2mm}
\renewcommand\thetable{2}
 \caption{SI-SNRi (dB) of downstream models with pre-trained utterance-level SID embeddings (higher is better).}
 \begin{tabular}{l|cccc|cccc}
 \hline
 {} & \multicolumn{4}{c|}{LibriSpeech} & \multicolumn{4}{c}{VCTK} \\
 \cline{2-9}
 {} & E3Net & Conv-TasNet & VoiceFilter & pDCCRN & E3Net & Conv-TasNet & VoiceFilter & pDCCRN \\
 \hline 
 FBANK & 11.3 & 13.3 & 8.8 & 9.3 & 9.4 & 11.2 & 6.9 & 7.7 \\
 d-vector & 11.5 & 13.2 & 9.2 & 10.0 & 9.1 & 11.9 & 7.1 & 8.2 \\
 e-vector & 10.3 & 12.4 & 8.8 & 9.7 & 8.5 & 11.1 & 6.9 & 7.9 \\
 \hline
 d-vector-FN & 9.7 & 11.3 & 8.8 & 9.5 & 7.7 & 10.2 & 7.1 & 7.7 \\
 e-vector-FN & 8.7 & 11.4 & 8.4 & 9.3 & 7.2 & 10.4 & 6.6 & 7.6 \\
 \hline
 d-vector-aug & 11.5 & 13.5 & 9.2 & 9.9 & 9.2 & 12.0 & 7.1 & 8.4 \\
 \hline
 \end{tabular}
\vspace{-2mm}
 \label{table:utt_SID}
\end{table*}

\begin{table}[t]
\vspace{-3mm}
\centering
\renewcommand\thetable{1}
  \caption{EER (\%) of the FBANK and SID systems (lower is better).}
\begin{tabular}{l|ccc}
\hline
{} & VoxCeleb1 & LibriSpeech & VCTK \\
\hline 
FBANK & 38.0 & 14.7 & 29.6 \\
d-vector & 5.3 & 3.9 & 4.4 \\
e-vector & 1.7 & 1.8 & 1.6 \\
\hline
d-vector-FN & 4.1 & 3.7 & 3.7 \\
e-vector-FN & 1.4 & 1.7 & 1.3 \\
\hline
d-vector-aug & 3.3 & 1.3 & 2.3 \\
\hline
\end{tabular}
\vspace{-4mm}
\label{table:eer}
\end{table}

\subsection{Downstream separation models}

Four popular separation models (or a subset) are used to test the performance of the utterance-level embeddings:
\begin{itemize}
\item \textbf{E3Net and Conv-TasNet}: These two belong to the waveform-based encoder-separator-decoder scheme~\cite{e3net, luo2019conv}. The encoder transforms each $L$\,ms mixture frame shifted by $S$\,ms into an $N$-dim latent space, which is projected down to $B$~dimensions by the bottleneck layer in the separator. The rest of the separator contains either $R$~LSTM networks in E3Net or temporal convolutional networks in Conv-TasNet. In E3Net, the enrollment features are concatenated with the encoder output, while in Conv-TasNet they are fused with each of the $X$ \mbox{1-D} convolutional blocks by a FiLM layer~\cite{perez2018film}. The decoder does the inverse transform. For E3Net, we set $L=20$, $S=10$, $N=2048$, $B=256$, and $R=4$. For Conv-TasNet, we set $L=10$, $S=5$, $N=1024$, $B=256$, $R=2$, and $X=8$. The training objective is the scale-invariant signal-to-noise ratio (SI-SNR)~\cite{le2019sdr}.

\item \textbf{VoiceFilter and pDCCRN}: These two are STFT-based models. We implemented the VoiceFilter in~\cite{vfilter}, which sends the mixture magnitude STFT through a \mbox{2-D} CNN stack followed by an LSTM and fully connected (FC) layers to learn a mask. pDCCRN~\cite{pdccrn} is a TSS adaptation of the unconditional DCCRN~\cite{dccrn}, a complex-valued U-Net consisting of a convolutional encoder-decoder and an LSTM in between. In both models, the upstream embeddings are concatenated with the LSTM input. We refer to the official DCCRN code~\cite{dccrn_code} and the ``student'' pDCCRN configuration as in~\cite{e3net}. Both models are trained with the power-law compressed phase-aware asymmetric loss defined in~\cite{pdccrn}.
\end{itemize}

\begin{figure}[t]
    \centering
    \includegraphics[scale=0.45]{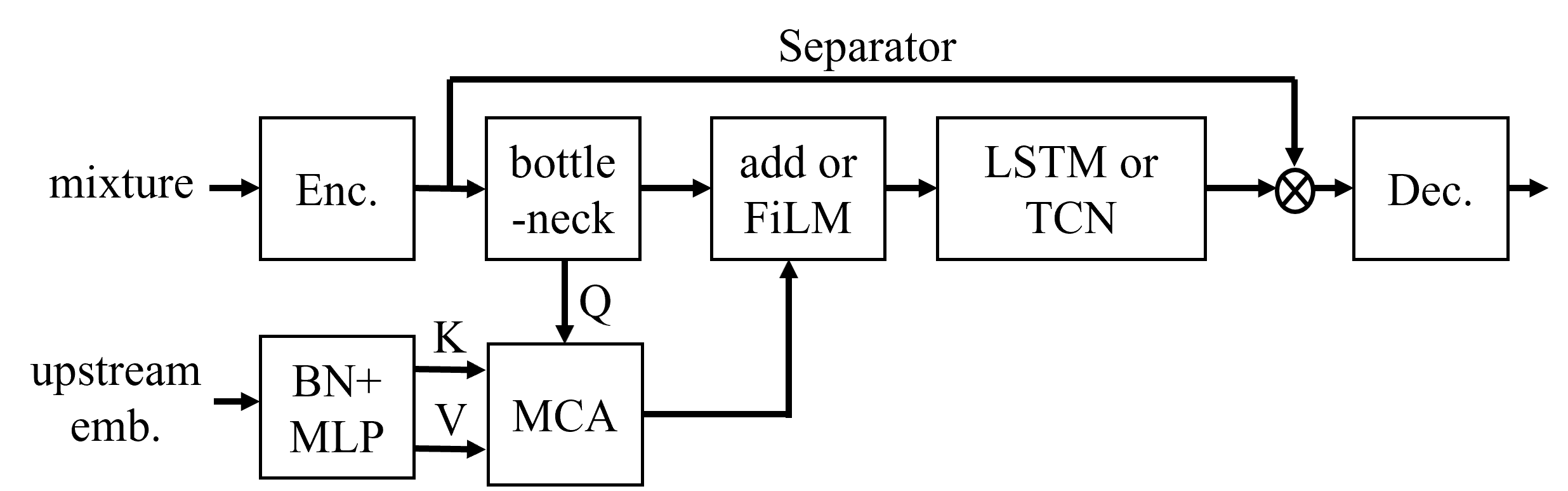}
    \vspace{-6mm}
    \caption{Downstream framework for frame-level embeddings.}
    \vspace{-2mm}
    \label{fig:framework}
    \vspace{-1mm}
\end{figure}

Before conditioning the downstream, we optionally center the embeddings with a \mbox{1-D} batch normalization (BN), and always map the outputs to 256-dim by a multi-layer perceptron (MLP), with 2 FC+PReLU layers, so that the downstream model size is invariant to the embeddings' size. The BN ensures convergence in some cases, and is removed if degradation is observed. We also did not observe a performance change that may bias the study due to the MLP.

E3Net and Conv-TasNet are adapted for testing the frame-level embeddings (Fig.~\ref{fig:framework}). A multi-head cross-attention (MCA) layer based on~\cite{all_you_need} aligns the mixture frames after the bottleneck layer (queries Q) with the BN+MLP-processed enrollment frames (keys K and values V). Inside the MCA, sinusoidal positional encoding~\cite{all_you_need} is added to Q, K, and V, and then a layer normalization is used before the QKV transforms. We use 4 heads and condition with simple addition for E3Net and with FiLM for Conv-TasNet. To obtain the scale and bias for FiLM, we use two parallel FC layers in the MCA to merge the multi-head outputs (the transforms inside MCA increase the downstream size). We trained an E3Net with comparable size but without attention and did not see improvements, which verifies the effectiveness of attention. All the downstream models are trained for 500\,k iterations using the Adam optimizer with a cosine learning rate scheduler. The peak learning rate is 0.0001 for E3Net, 0.0005 for Conv-TasNet and VoiceFilter, and 0.001 for pDCCRN.

\subsection{Data, evaluation metrics, and SID performance}
\label{subsec:data}
The downstream training set is created by online mixing the 960-hour LibriSpeech training set (2,338 speakers). Each mixture is the non-scaled sum of two utterances from a target and an interfering speaker, respectively. Another utterance from the target speaker is used for enrollment. The interfering speech is either segmented or repeated to the target speech length.

As mentioned, we are interested in a cross-dataset evaluation. 
To evaluate the in-dataset performance, we use the VoiceFilter LibriSpeech dev-clean+dev-other list~\cite{vfilter_list} (5,567~samples from 73~unseen speakers). For the cross-dataset generalization test, we use the VoiceFilter VCTK test list~\cite{vfilter_list} (4,212~samples from 10 speakers). Note that VCTK is only used for evaluation. To make the task more challenging, we employ VCTK 0.92~\cite{vctk} and use two different microphones for enrollment and mixture, respectively. We trim silences in VCTK before mixing, and make 20\,s enrollment utterances by concatenating utterances from the same VCTK target speaker. 

All the data we use is 16\,kHz. The test set mixture SI-SNR before TSS processing is $-0.3$\,dB for LibriSpeech and $-0.5$\,dB for VCTK. For conciseness, we here only report the SI-SNR improvement after TSS (SI-SNRi). We verified that other metrics SDRi, PESQ, and STOI showed the same trends as SI-SNRi for all results.

Before focusing on TSS, we first examine the EERs of the SID systems (Table~\ref{table:eer}) using 50\,k trial pairs formed by the same TSS test speakers in LibriSpeech and VCTK. The utterances in each VCTK pair use different microphones. For completeness, we also show the EERs computed directly upon the cosine distance of the utterance-level FBANK embedding (which is not common practice). The high EERs imply that FBANK is not as speaker-discriminative as other embeddings. Next, for systems without data augmentation, e-vectors yield lower EERs than d-vectors. Consistent with previous work~\cite{zhao2020improving, zheng2006comparative}, FN reduces EERs in all datasets. Data augmentation (d-vector-aug vs.\ d-vector) greatly improves performance, especially on LibriSpeech due to similar data used for augmentation.

\vspace{-2mm}

\section{Experiments and Discussion}
\label{sec:experiments}
\vspace{-1mm}
\subsection{Pre-trained utterance-level SID embeddings}
\label{subsec:sid}

We now quantify and discuss the answers to the questions formulated in Sec.~\ref{sec:introduction}. We first focus on the results of Table~\ref{table:utt_SID}:

\begin{itemize}
\item \textbf{How does each embedding compare with FBANK}? Remarkably, the simple FBANK performs comparably with other SID embeddings and, in some cases, even better than e-vector, the state-of-the-art SID model. This suggests that, although FBANK is not as speaker-discriminative as SID embeddings, it provides enough useful enrollment information for the downstream TSS models. 

\item \textbf{Does a low EER mean better separation}? Though a low EER is often preferred for TSS~\cite{two-stage, speaker-aware, percepnet}, we see that this does not necessarily imply higher TSS quality. The d-vector consistently attains a higher SI-SNRi than the e-vector, which reports lower EERs. We hypothesize that a more capable discriminative SID model may remove more SID-irrelevant information to achieve a lower EER, but the lost information might be actually needed by TSS.

\item \textbf{Does FN in SID improve TSS}? Although FN contributes to better EERs by normalizing the stationary recording characteristics, we see that it consistently hurts TSS, with the largest SI-SNRi drop being $-$1.9\,dB. Large degradation is also observed on the 2-microphone VCTK test set. This trend shows that the information in the FBANK mean subtracted from the input to the upstream SID systems is important for the downstream TSS task.

\begin{table}[t]
\vspace{-3mm}
\centering
\renewcommand\thetable{3}
  \caption{LibriSpeech SI-SNRi (dB) when test enrollment is noisy. Training enrollment can be clean (mismatched) or noisy (matched).}
\begin{tabular}{l|cc|cc}
\hline
{} & \multicolumn{2}{c|}{Mismatched} & \multicolumn{2}{c}{Matched} \\
\cline{2-5}
{} & E3Net & Conv-TasNet & E3Net & Conv-TasNet \\
\hline 
FBANK & 9.9 & 11.1 & 10.5 & 12.1 \\
d-vec-aug & 10.7 & 12.3 & 11.0 & 12.7 \\
\hline
\end{tabular}
\vspace{-3mm}
\label{table:noisy}
\end{table}

\begin{table*}[t]
\centering
\vspace{-5mm}
\renewcommand\thetable{4}
 \caption{SI-SNRi (dB) of downstream models with pre-trained utterance-level SSL features (higher is better). Also show FBANK and d-vector for comparison. Conv-TasNet is conditioned on upstream embeddings through the bottleneck layer.}
 \begin{tabular}{l|ccc|ccc}
 \hline
 {} & \multicolumn{3}{c|}{LibriSpeech} & \multicolumn{3}{c}{VCTK} \\
 \cline{2-7}
 {} & E3Net & Conv-TasNet & pDCCRN & E3Net & Conv-TasNet & pDCCRN \\
 \hline 
 FBANK & 11.3 & 13.6 & 9.3 & 9.4 & 11.7 & 7.7 \\
 d-vector & 11.5 & 13.3 & 10.0 & 9.1 & 11.1 & 8.2 \\
 \hline
 PASE+ & 11.6 & 13.8 & 9.6 & 8.8 & 10.6 & 7.5 \\
 HuBERT & 11.8 & 14.1 & 10.0 & 8.7 & 10.2 & 7.3 \\
 \hline

 \end{tabular}
\vspace{-6mm}
 \label{table:utt_ssl}
\end{table*}

\item \textbf{Does data augmentation in SID improve TSS}? Relative to d-vector, d-vector-aug is trained with (i) more speakers and (ii) artificial distortions (Sec.~\ref{sec:upstream}). First, we test the effects of (i) based on Table~\ref{table:utt_SID} under a clean enrollment condition. We do not observe gains in going from d-vector (7\,k) to d-vector-aug (18\,k speakers). We hypothesize that SID may not extract the key information for TSS despite of more training speakers. Next, we test the effects of (ii) on the robustness of d-vector-aug based on Table~\ref{table:noisy} under a noisy enrollment condition. Since the d-vector training set (VoxCeleb) already contains some noisy data (before augmentation), we instead use FBANK as the baseline. We corrupt the downstream LibriSpeech training and test enrollment data with noise and reverb from~\cite{dns-challenge} while keeping mixtures and targets clean. If we reuse the downstream models trained with the clean enrollment (mismatched training/testing condition), d-vector-aug shows larger advantage. Nonetheless, if we re-train the downstream models with the corrupted enrollment data (the matched case), we see that the gap shrinks. Thus, we infer that the FBANK can be made comparably robust by augmenting the downstream enrollment. 
\end{itemize}

\vspace{-3mm}
\subsection{Pre-trained utterance-level SSL embeddings}
\label{subsec:ssl}

Previous research shows that SSL outperforms FBANK for encoding speech information~\cite{superb}, but SSL has not been explored for TSS enrollment. Table~\ref{table:utt_ssl} shows the results for PASE+ and HuBERT-Large, together with FBANK and d-vector (for Conv-TasNet to be comparable with Sec.~\ref{subsec:attention}, FiLM conditioning is applied to the bottleneck layer). We will now first answer the questions using LibriSpeech and then study generalization to VCTK. 

\begin{itemize}
\item \textbf{How does SSL compare with FBANK? Does a more powerful SSL model yield better TSS}? On LibriSpeech, SSL outperforms FBANK by a small amount. HuBERT-Large, the more powerful model, provides some improvement, but at the cost of a very large and complicated upstream structure (316\,M parameters), large training data (60\,k hours), and GPU requirements.

\item \textbf{Which one is better, SSL or SID embeddings?} On LibriSpeech, the largest SSL improvement over d-vector is 0.8\,dB (HuBERT Conv-TasNet) but, in other cases, the difference is very small. We will see in Sec.~\ref{subsec:attention} that the utterance mean underutilizes the frame-level information encoded by SSL, preventing larger improvements over d-vector.


\item \textbf{Generalization to VCTK?} Out of the three datasets: (i) upstream training, (ii) downstream training, and (iii) downstream evaluation, previous SSL works often report large improvements relative to FBANK when at least (ii) and (iii) are partitions of the same dataset~\cite{superb} (in-dataset training/fine-tuning and evaluation). Nonetheless, as mentioned, we are here interested in the case where (iii) is different from (i) and (ii) (cross-dataset evaluation). In that situation, we observe that SSL is worse than FBANK, which suggests that the SSL embeddings may overfit to the SSL objective or the upstream data. This overfitting could partially contribute to the gain on LibriSpeech, since (i), (ii), and (iii) are partitions of the same dataset (LibriVox). We also observe that the larger and more powerful HuBERT overfits more, despite of the 60\,k-hour Libri-Light upstream training set. One potential reason for these results could be lack of data diversity for the upstream training~\cite{towards_generalization} or lack of regularization in the embedding space~\cite{simplicial_emb}.

\end{itemize}

\vspace{-3mm}

\subsection{Pre-trained frame-level embeddings with attention}
\label{subsec:attention}
We now summarize the results based on pre-trained frame-level embeddings (Table~\ref{table:attention}). We note that the Conv-TasNet attention on VCTK degrades with respect to the FBANK and d-vector utterance-level results (compare with Table~\ref{table:utt_ssl}). This requires further investigation. However, since attention brings improvements in all other cases (14 out of 16 evaluations), we still rely on this test bed but will exclude those two VCTK results from our observations. 

\begin{table}[t]
\centering
\renewcommand\thetable{5}
  \caption{SI-SNRi (dB) using pre-trained frame-level embeddings with attention.}
\begin{tabular}{l|cc|cc}
\hline
{} & \multicolumn{2}{c|}{LibriSpeech} & \multicolumn{2}{c}{VCTK} \\
 \cline{2-5}
{} & E3Net & Conv-TasNet & E3Net & Conv-TasNet \\
\hline 
FBANK & 12.0 & 14.2 & 10.1 & 11.3 \\
d-vector & 11.6 & 13.5 & 9.1 & 10.6 \\
\hline
PASE+ & 12.1 & 14.4 & 9.9 & 11.3 \\
HuBERT & 12.3 & 14.6 & 9.9 & 11.0 \\
\hline
\end{tabular}
\vspace{-3mm}
\label{table:attention}
\end{table}

\begin{itemize}
\item \textbf{Which one is better, FBANK, SSL, or SID embeddings}? First, we look at SSL and SID. Compared with the utterance-level results in Table~\ref{table:utt_ssl}, both SSL features benefit from frame-level information, but the frame-level d-vector does not. Frame-level information leads to larger advantages for SSL, with the largest improvement being 1.1\,dB. Since the d-vector training objective is still cross-entropy, removing the temporal pooling cannot compensate for the lost information. Next, comparing FBANK and the rest, SSL is only slightly better than FBANK on LibriSpeech, but that again does not generalize to the VCTK test set. FBANK also outperforms d-vector, with the largest gap being 1\,dB, showing again the benefit of preserving complete enrollment information.     
\end{itemize}

\vspace{-3mm}
\subsection{Fine-tuned frame-level embeddings}

\begin{table}[t]
\centering
\renewcommand\thetable{6}
  \caption{E3Net SI-SNRi (dB) using fine-tuned frame-level embeddings with attention.}
\begin{tabular}{l|c|c}
\hline
{} & LibriSpeech & VCTK \\
\hline 
FBANK & 12.0 & 10.1 \\
\hline
Frozen d-vector & 11.6 & 9.1 \\
Fine-tuned d-vector & 11.8 & 9.2 \\
\hline
Frozen PASE+ & 12.1 & 9.9 \\
Fine-tuned PASE+ & 12.6 & 9.6 \\
\hline
\end{tabular}
\vspace{-4mm}
\label{table:finetune}
\end{table}

We fine-tune the pre-trained PASE+ and d-vector models with the E3Net on LibriSpeech, and evaluate the performance on both test sets. We answer the following two questions based on Table \ref{table:finetune}.

\begin{itemize}
\item \textbf{Do fine-tuned embeddings outperform the frozen ones}? Fine-tuning PASE+ brings a 0.5\,dB gain over the frozen one on the LibriSpeech test set. This agrees with previous works~\cite{speaker-aware, time-speakerbeam}. However, the improvement does not generalize to the VCTK test set. There, the fine-tuned PASE+ is worse, indicating that the increased model size might cause some overfitting. Note that both works~\cite{speaker-aware, time-speakerbeam} lack the cross-dataset evaluation.   

\item \textbf{Which one is more effective, fine-tuning SSL or SID}? We do not observe meaningful gains from fine-tuning the d-vector. However, we only use the separation loss, whereas ~\cite{speaker-aware, time-speakerbeam} show the benefits of moving the upstream SID loss to the downstream to compensate for the separation loss. This is to be further investigated.   
\end{itemize}

\vspace{-2mm}    

\section{Conclusion}
\label{sec:conclusion}

To summarize, TSS enrollment embeddings need to (i) preserve the integrity of the relevant information, and (ii) generalize to different test sets. On (i), SID embeddings could cause degradation through previously overlooked factors, such as a sub-optimal metric (EER), training objective (cross-entropy), or pre-processing (FN). The lost information could make data augmentation and the attention algorithm less effective. On (ii), we see that both the pre-trained and fine-tuned SSL features do not generalize well to the VCTK test set in the cross-dataset study. Interestingly, we find that the simple yet overlooked FBANK meets both (i) and (ii), providing both competitive separation and generalization performance. That is not to say that FBANK may be the ultimate solution, but to provide more insight, and to encourage the development of better features that can clearly outperform FBANK in both separation and generalization terms.

\bibliographystyle{IEEEbib}
\bibliography{z_bibliography}

\begin{thebibliography}{10}

\bibitem{vfilter}
Q.~Wang~et al.,
\newblock ``{VoiceFilter: Targeted Voice Separation by Speaker-Conditioned
  Spectrogram Masking},''
\newblock in {\em INTERSPEECH}, 2019, pp. 2728--2732.

\bibitem{percepnet}
R.~Giri et~al.,
\newblock ``{Personalized PercepNet: Real-Time, Low-Complexity Target Voice
  Separation and Enhancement},''
\newblock in {\em INTERSPEECH}, 2021, pp. 1124--1128.

\bibitem{e3net}
M.~Thakker~et al.,
\newblock ``{Fast Real-time Personalized Speech Enhancement: End-to-End
  Enhancement Network (E3Net) and Knowledge Distillation},''
\newblock {\em arXiv preprint arXiv:2204.00771}, 2022.

\bibitem{pdccrn}
S.~E. Eskimez~et al.,
\newblock ``{Personalized Speech Enhancement: New Models and Comprehensive
  Evaluation},''
\newblock in {\em ICASSP}, 2022, pp. 356--360.

\bibitem{two-stage}
Y.~Ju~et al.,
\newblock ``{TEA-PSE: Tencent-Ethereal-Audio-Lab Personalized Speech
  Enhancement System for ICASSP 2022 DNS Challenge},''
\newblock in {\em ICASSP}, 2022, pp. 9291--9295.

\bibitem{atss-net}
T.~Li~et al.,
\newblock ``{Atss-Net: Target Speaker Separation via Attention-based Neural
  Network},''
\newblock in {\em INTERSPEECH}, 2020, pp. 1411--1415.

\bibitem{one-enhance-all}
H.~Taherian~et al.,
\newblock ``{One Model to Enhance Them All: Array Geometry Agnostic
  Multi-Channel Personalized Speech Enhancement},''
\newblock in {\em ICASSP}, 2022, pp. 271--275.

\bibitem{speaker-inventory}
X.~Xiao~et al.,
\newblock ``{Single-channel Speech Extraction Using Speaker Inventory and
  Attention Network},''
\newblock in {\em ICASSP}, 2019, pp. 86--90.

\bibitem{attention-scaling}
J.~Han~et al.,
\newblock ``{Attention-Based Scaling Adaptation for Target Speech
  Extraction},''
\newblock in {\em ASRU}, 2021, pp. 658--662.

\bibitem{speaker-aware}
X.~Ji~et al.,
\newblock ``{Speaker-Aware Target Speaker Enhancement by Jointly Learning with
  Speaker Embedding Extraction},''
\newblock in {\em ICASSP}, 2020, pp. 7294--7298.

\bibitem{time-speakerbeam}
M.~Delcroix~et al.,
\newblock ``{Improving Speaker Discrimination of Target Speech Extraction with
  Time-Domain SpeakerBeam},''
\newblock in {\em ICASSP}, 2020, pp. 691--695.

\bibitem{pascual2019learning}
M.~Ravanelli~et al.,
\newblock ``{Multi-Task Self-Supervised Learning for Robust Speech
  Recognition},''
\newblock in {\em ICASSP}, 2020, pp. 6989--6993.

\bibitem{wav2vec2}
A.~Baevski~et al.,
\newblock ``{wav2vec 2.0: A Framework for Self-Supervised Learning of Speech
  Representations},''
\newblock {\em NeurIPS 2020}, vol. 33, pp. 12449--12460.

\bibitem{hubert}
W.~Hsu~et al.,
\newblock ``{HuBERT: Self-Supervised Speech Representation Learning by Masked
  Prediction of Hidden Units},''
\newblock {\em IEEE/ACM Transactions on Audio, Speech, and Language
  Processing}, vol. 29, pp. 3451--3460, 2021.

\bibitem{ssl-separation}
Z.~Huang~et al.,
\newblock ``{Investigating Self-Supervised Learning for Speech Enhancement and
  Separation},''
\newblock in {\em ICASSP}, 2022, pp. 6837--6841.

\bibitem{kadiouglu2020empirical}
B.~Kad{\i}o{\u{g}}lu~et al.,
\newblock ``{An Empirical Study of Conv-TasNet},''
\newblock in {\em ICASSP}, 2020, pp. 7264--7268.

\bibitem{zeinali2019but}
H.~Zeinali~et al.,
\newblock ``{BUT System Description to VoxCeleb Speaker Recognition Challenge
  2019},''
\newblock {\em arXiv preprint arXiv:1910.12592}, 2019.

\bibitem{ecapa-tdnn}
B.~Desplanques, J.~Thienpondt, and K.~Demuynck,
\newblock ``{ECAPA-TDNN: Emphasized Channel Attention, Propagation and
  Aggregation in TDNN Based Speaker Verification},''
\newblock in {\em INTERSPEECH}, 2020, pp. 3830--3834.

\bibitem{xvector}
D.~Snyder~et al.,
\newblock ``{X-Vectors: Robust DNN Embeddings for Speaker Recognition},''
\newblock in {\em ICASSP}, 2018, pp. 5329--5333.

\bibitem{zhao2020improving}
Y.~Zhao~et al.,
\newblock ``{Improving Deep CNN Networks with Long Temporal Context for
  Text-Independent Speaker Verification},''
\newblock in {\em ICASSP}, 2020, pp. 6834--6838.

\bibitem{zheng2006comparative}
R.~Zheng, S.~Zhang, and B.~Xu,
\newblock ``{A Comparative Study of Feature and Score Normalization for Speaker
  Verification},''
\newblock in {\em International conference on biometrics}, 2006, pp. 531--538.

\bibitem{voxceleb1}
A.~Nagrani, J.~S. Chung, and A.~Zisserman,
\newblock ``{VoxCeleb: A Large-Scale Speaker Identification Dataset},''
\newblock in {\em INTERSPEECH}, 2017, pp. 2616--2620.

\bibitem{voxceleb2}
J.~S. Chung, A.~Nagrani, and A.~Zisserman,
\newblock ``{VoxCeleb2: Deep Speaker Recognition},''
\newblock in {\em INTERSPEECH}, 2018, pp. 1086--1090.

\bibitem{dvector}
L.~Wan~et al.,
\newblock ``{Generalized End-to-End Loss for Speaker Verification},''
\newblock in {\em ICASSP}, 2018, pp. 4879--4883.

\bibitem{specaugment}
D.~S.~Park et~al.,
\newblock ``{SpecAugment: A Simple Data Augmentation Method for Automatic
  Speech Recognition},''
\newblock in {\em INTERSPEECH}, 2019, pp. 2613--2617.

\bibitem{dns-challenge}
H.~Dubey~et al.,
\newblock ``{ICASSP 2022 Deep Noise Suppression Challenge},''
\newblock in {\em ICASSP}, 2022, pp. 9271--9275.

\bibitem{arcface}
J.~Deng~et al.,
\newblock ``{ArcFace: Additive Angular Margin Loss for Deep Face
  Recognition},''
\newblock in {\em CVPR}, 2019, pp. 4685--4694.

\bibitem{speechbrain}
{\em {\textnormal{The SpeechBrain ECAPA-TDNN recipe can be retrieved from:
  ``https://github.com/speechbrain/speechbrain"} }}.

\bibitem{librispeech}
V.~Panayotov~et al.,
\newblock ``{LibriSpeech: An ASR Corpus Based on Public Domain Audio Books},''
\newblock in {\em ICASSP}, 2015, pp. 5206--5210.

\bibitem{superb}
S.~Yang et~al.,
\newblock ``{SUPERB: Speech processing Universal PERformance Benchmark},''
\newblock in {\em INTERSPEECH}, 2021, pp. 1194--1198.

\bibitem{libri-light}
J.~Kahn~et al.,
\newblock ``{Libri-Light: A Benchmark for ASR with Limited or No
  Supervision},''
\newblock in {\em ICASSP}, 2020, pp. 7669--7673.

\bibitem{luo2019conv}
Y.~Luo and N.~Mesgarani,
\newblock ``{Conv-TasNet: Surpassing Ideal Time–Frequency Magnitude Masking
  for Speech Separation},''
\newblock {\em IEEE/ACM Transactions on Audio, Speech, and Language
  Processing}, vol. 27, no. 8, pp. 1256--1266, 2019.

\bibitem{perez2018film}
E.~Perez~et al.,
\newblock ``{FiLM: Visual Reasoning with a General Conditioning Layer},''
\newblock in {\em AAAI}, 2018, vol.~32.

\bibitem{le2019sdr}
J.~Le~Roux~et al.,
\newblock ``{SDR -- Half-Baked or Well Done?},''
\newblock in {\em ICASSP}, 2019, pp. 626--630.

\bibitem{dccrn}
Y.~Hu et~al.,
\newblock ``{DCCRN: Deep Complex Convolution Recurrent Network for Phase-Aware
  Speech Enhancement},''
\newblock in {\em INTERSPEECH}, 2020, pp. 2472--2476.

\bibitem{dccrn_code}
{\em {\textnormal{The DCCRN code used by this work can be retrieved from:
  ``https://github.com/huyanxin/DeepComplexCRN"} }}.

\bibitem{all_you_need}
A.~Vaswani~et al.,
\newblock ``{Attention is All You Need},''
\newblock {\em NeurIPS 2017}, vol. 30.

\bibitem{vfilter_list}
{\em {\textnormal{The VoiceFilter test lists used in this work can be retrieved
  from: ``https://google.github.io/speaker-id/publications/VoiceFilter/"} }}.

\bibitem{vctk}
J.~Yamagishi, C.~Veaux, and K.~MacDonald,
\newblock ``{CSTR VCTK Corpus: English Multi-speaker Corpus for CSTR Voice
  Cloning Toolkit} (version 0.92),'' 2019.

\bibitem{towards_generalization}
W.~Huang, M.~Yi, and X.~Zhao,
\newblock ``{Towards the Generalization of Contrastive Self-Supervised
  Learning},''
\newblock {\em arXiv preprint arXiv:2111.00743}, 2021.

\bibitem{simplicial_emb}
S.~Lavoie~et al.,
\newblock ``{Simplicial Embeddings in Self-Supervised Learning and Downstream
  Classification},''
\newblock {\em arXiv preprint arXiv:2204.00616}, 2022.

\end{thebibliography}

\end{document}